\RequirePackage{amsmath}
\documentclass[nofootinbib,floatfix,pra,aps,twocolumn,10pt]{revtex4-1}

\usepackage{color}

\usepackage[utf8]{inputenc}
\usepackage{hyperref}
\let\oldvec\vec
\usepackage{amsmath}
\let\vec\oldvec
\usepackage{bbold}
\usepackage{amssymb}
\usepackage{graphicx}
\usepackage{mciteplus}
\usepackage[absolute]{textpos}

\newcommand{\beq}{\begin{equation}}
\newcommand{\eeq}{\end{equation}}
\newcommand{\bea}{\begin{eqnarray}}
\newcommand{\eea}{\end{eqnarray}}

\begin{document}

\title{Thermodynamics of spin-orbit coupled bosons in two dimensions \texorpdfstring{\\}{ } from complex Langevin}
\author{Felipe Attanasio}
\email{pyfelipe@uw.edu}
\affiliation{Department of Physics, University of Washington, Box 351560, Seattle, Washington 98195-1560, USA}
\author{Joaqu\'in E. Drut}
\email{drut@email.unc.edu}
\affiliation{Department of Physics and Astronomy, University of North Carolina, Chapel Hill, North Carolina 27599-3255, USA}
\begin{abstract}
We investigate the thermal properties of interacting spin-orbit coupled bosons with contact interactions in two spatial dimensions.
To that end, we implement the complex Langevin method, motivated by the appearance of a sign problem, on a square lattice with periodic boundary conditions.
We calculate the density equation of state non-perturbatively in a range of spin-orbit couplings and chemical potentials.
Our results show that mean-field solutions tend to underestimate the average density, especially for stronger values of the spin-orbit coupling.
Additionally, the finite nature of the simulation volume induces the formation of pseudo-condensates.
These have been observed to be destroyed by the spin-orbit interactions.
\begin{textblock}{20}(11.4,0.3)
	\begin{flushleft}
		NT@UW-19-10, INT-PUB-19-035
  \end{flushleft}
 \end{textblock}%
\end{abstract} 
\maketitle
%

\section{Introduction}

The experimental realization of ultracold atomic systems with spin-orbit coupling (SOC), nearly a decade ago~\cite{lin_spinorbit-coupled_2011,wang_spin-orbit_2012,cheuk_spin-injection_2012}, opened an
exciting new set of directions for the exploration of the properties of matter in extreme (yet highly controllable) conditions. 
The SOC, which couples the  atomic pseudo-spin 
(which itself denotes the particle species or `flavor') to momentum, is realized as the coupling of conventional
nonrelativistic neutral particles to a synthetic non-abelian background gauge field~\cite{lin_synthetic_2009,lin_synthetic_2011,Dalibard:2010ph}. 
Such a construction has potential applications for the exploration of a variety of physical situations (including Rashba-~\cite{ISI:A1960WT33500031} and Dresselhaus-type~\cite{PhysRev.100.580} couplings).
Notably, these systems have been under intense scrutiny in recent years as they may be used experimentally to 
realize exotic phases of matter such as supersolids~\cite{wu_realization_2016,leonard_supersolid_2017,2017Natur.543...91L}, superfluids with a crystalline structure, and study exotic topological properties~\cite{2016NatPh..12..540H,PhysRevLett.117.235304,2019arXiv190708637V}.

More generally, ultracold bosons subject to SOC are known, at mean field level, to exhibit stripe or plane wave 
phases~\cite{PhysRevLett.105.160403, Zhai:2014gna}: the ground state wave function in the former is composed of two plane 
waves propagating in opposite directions, leading to an interference pattern, while on the latter it has only one plane wave.
Many recent studies have investigated mean field properties of different types of SOC~\cite{2017PhRvA..95e1601K}, and also 
first order (one loop) quantum corrections~\cite{2018PhRvL.120n0403L, 2019arXiv190308182L}.
Recent theoretical studies of spin-orbit coupling include its effect combined with finite angular momentum~\cite{2017PhRvL.118n5302S} and a harmonic trap~\cite{2019arXiv190713355M}.

In this work we characterize, in a non-perturbative fashion, some of the basic thermodynamic observables of spin-orbit coupled 
bosons with quartic interactions. Using the complex Langevin method, we perform a study of such a non-relativistic Bose gas on a spacetime lattice, 
determining the density equation of state, pressure and pseudo-condensate fraction.
Notably, the lattice formulation compactifies the spin-orbit interaction, which appears as a constant background non-abelian gauge field.
This allows for an easy study of the limit of very large SOC compared to the momentum.

\section{Model and lattice formulation}

We consider a system of $(2+1)$-dimensional non-relativistic bosons with two hyperfine (pseudo-spin) states, denoted by $\uparrow$ 
and $\downarrow$, in Euclidean spacetime.
They are subject to a Rashba-Dresselhaus spin-orbit coupling (SOC) and density-density contact interactions. More specifically, we will use the Euclidean action
\beq
S = \int d^2x d\tau \, \mathcal L,
\eeq
where $\mathcal L = \mathcal{L}_0  + \mathcal{L}_{\mathrm{int}}$,
\begin{align}
\mathcal{L}_0 =& \Phi^\dagger ( \partial_\tau - \mu ) \Phi \nonumber\\
& -  \Phi^\dagger  \frac{(-i \openone\partial_x-\kappa_x \sigma_x)^2 + (-i \openone\partial_y-\kappa_y \sigma_y)^2}{2m}  \Phi  \,,
\end{align}
is the free part, and
\begin{align}
\mathcal{L}_{\mathrm{int}} = 
\frac{\lambda}{4} \left[\left(\phi^\dagger_\uparrow \phi^{}_\uparrow \right)^2 + \left(\phi^\dagger_\downarrow \phi^{}_\downarrow \right)^2\right] 
+ \frac{g}{2} \left(\phi^\dagger_\uparrow \phi^{}_\uparrow \phi^\dagger_\downarrow \phi^{}_\downarrow \right) \,,
\end{align}
is the interacting part. In the above equations, $\Phi = [\phi_\uparrow \quad \phi_\downarrow]^T$, $\mu$ is the chemical potential
for both species, $m$ is the mass, $\lambda$ and $g$ are the intra- and inter-species couplings, and $\kappa_x$ and $\kappa_y$ 
characterize the spin-orbit interaction coupled to the $\sigma_x$ and $\sigma_y$ Pauli matrices.
Note that, by virtue of the SOC, neither particle number is individually conserved, but the total particle number is,
such that $\mu$ is a sensible chemical potential. 
 
In order to perform our lattice studies, we discretise the system in a hypercubic lattice of size $N_x^2 \times N_\tau$ and spacings $a$ and $a_\tau$ in the spatial and temporal directions, respectively.
The temperature is given by $T=\beta^{-1}=1/a_\tau N_\tau$.
We assume periodic boundary conditions in all directions.
The SOC enters the action in the same way as a background SU($2$) gauge field, similar to a minimal coupling.
On the lattice, the SOC is treated as a background non-abelian gauge field and it is discretized in the same way.
The discretized version of the action, with the fields and couplings being rescaled by appropriate powers of the lattice spacing, is given by
\begin{align}
S = &\, \xi \sum_{\vec{x},\tau} \left\{ \Phi^\dagger_{(\vec{x},\tau)} \left (\Phi_{(\vec{x},\tau)} - e^{\xi \mu} \Phi_{(\vec{x},\tau-a_\tau)}\right) \xi^{-1} \right. \nonumber\\
&- \frac{1}{2m} \sum_j \Phi^\dagger_{(\vec{x},\tau)} \left[ v_j \Phi_{(\vec{x}+a \hat{j},\tau)} + v_j^\dagger \Phi_{(\vec{x}-a \hat{j},\tau)} - 2\Phi_{(\vec{x},\tau)} \right] \nonumber \\
&+ \frac{\lambda+g}{8} \left (\Phi^\dagger_{(\vec{x},\tau)} \Phi_{(\vec{x},\tau-a_\tau)} \right)^2 \nonumber\\
&+ \left. \frac{\lambda-g}{8} \left (\Phi^\dagger_{(\vec{x},\tau)} \sigma_z \Phi_{(\vec{x},\tau-a_\tau)} \right)^2 \right\}\,,
\end{align}
where $\hat{j}$ represents a unit vector in the $j$-th direction, $v_j = e^{-i \kappa_j \sigma_j}$, and $\xi = a_\tau / a$ is the lattice spacing anisotropy factor. The contact interactions have been regularized in the same way as the number density operator.
This formulation displays explicitly the global SU($2$) flavor symmetry of the contact interactions when $\lambda=g$, and the conservation of the total particle number by all interactions due to the global U($1$) symmetry.

In this work, we study the interplay between the self-coupling $\lambda$ and the spin-orbit couplings $\kappa_x$ and $\kappa_y$.
The coupling $g$ between different pseudo-spins is left for a future publication.

\subsection{Exact solution in the quadratic case}

When the quartic terms are not present, the action (hereafter referred to as $S_{\mathrm{free}}$) can be written 
in momentum space as
\beq
S_{\mathrm{free}} = \sum_{\vec{p},\omega} \tilde{\Phi}^\dagger_{(\vec{p},\omega)} M(\vec{p},\omega,\mu,\vec{\kappa}) \tilde{\Phi}_{(\vec{p},\omega)} \,,
\eeq
with
\beq
M=\begin{bmatrix} T + v_x^+ + v_x^- + v_y^+ + v_y^- & v_x^+ - v_x^- - i(v_y^+ - v_y^-) \\
	v_x^+ - v_x^- + i(v_y^+ - v_y^-) & T+ v_x^+ + v_x^- + v_y^+ + v_y^-\end{bmatrix}\,,
\eeq
where $v_j^\pm = \xi \sin^2((p_j \pm \kappa_j)/2)/m$, and $T=(1-e^{\xi\mu} e^{i \omega_q})$.
The lattice momenta and Matsubara frequencies are given by $p_j = 2 \pi k_j / N_x$, with $k_j = 0, \ldots, N_x-1$, and $\omega_q = 2\pi q/N_\tau$ with $q= 0, \ldots, N_\tau-1$, respectively. The tildes represent Fourier transforms.
The matrix $M$ can be diagonalized via a change to the helicity basis:
\beq
\begin{bmatrix} \varphi_+ \\ \varphi_- \end{bmatrix} =
\begin{bmatrix}
	\frac{(v^+_x - v^-_x) + i(v^+_y - v^-_y)}{\sqrt{2[(v^+_x - v^-_x)^2 + (v^+_y - v^-_y)^2]}} & \frac{1}{\sqrt{2}} \\
	-\frac{(v^+_x - v^-_x) + i(v^+_y - v^-_y)}{\sqrt{2[(v^+_x - v^-_x)^2 + (v^+_y - v^-_y)^2]}} & \frac{1}{\sqrt{2}}
\end{bmatrix}
\begin{bmatrix} \phi_\uparrow \\ \phi_\downarrow \end{bmatrix} \,,
\eeq
with eigenvalues
\begin{align}
\lambda_\pm(\vec{p},\omega,\mu,\vec{\kappa}) &= T + v_x^+ + v_x^- + v_y^+ + v_y^- \nonumber\\
&\pm \sqrt{(v_x^+ - v_x^-)^2 + (v_y^+ - v_y^-)^2}\,.
\end{align}
leading to the grand thermodynamic potential $\Omega$ via
\begin{align}
&\beta \Omega_{\mathrm{free}}(\mu,\vec{\kappa}) = -\ln \mathcal Z_{\mathrm{free}} \nonumber\\
&\quad = \sum_{\vec{p},\omega} [ \ln \lambda_+(\vec{p},\omega,\mu,\vec{\kappa}) + \ln \lambda_-(\vec{p},\omega,\mu,\vec{\kappa}) ] \,.
\end{align}
It is clear from the above equation that, since $p_x$ and $p_y$ run over the same interval, the free energy is symmetric under 
$\kappa_x \leftrightarrow \kappa_y$.
Moreover, the partition function on the lattice is an even, periodic function of the spin-orbit couplings, with period $\pi$ (in lattice units).
This periodicity is a lattice artifact that disappears in the continuum limit.

The average density of the noninteracting case can be obtained by differentiation with respect to $\beta \mu$:
\beq
\langle n \rangle = \frac{1}{V}\frac{\partial \ln \mathcal Z}{\partial (\beta \mu)} = 
\sum_{s = \pm}\sum_{\vec{p},q} \frac{1}{1 - X_{s}(\vec p, \mu) e^{-i \omega_q}},
\eeq
where
\beq
X_s(\vec p, \mu) = e^{- \xi \mu} (1 + \lambda_s - T).
\eeq
It can be shown that the sum over the Matsubara frequencies can be carried out, which yields
\beq
\langle n \rangle  = \frac{1}{V} \sum_{s=\pm} \sum_{\vec{p}} \frac{1}{[X_s(\vec p, \mu)]^{N_\tau} - 1} \, .
\eeq
In the $\xi \to 0$ limit, 
\beq
[X_s(\vec p, \mu)]^{N_\tau} \to e^{\beta (\epsilon_s(\vec p, \vec \kappa) - \mu)}
\eeq
where, in the continuum limit,
\begin{align}
\epsilon_\pm(\vec p, \vec \kappa) &=  \frac{|\vec{p}|^2 + |\vec{\kappa}|^2}{2m}\nonumber\\
&\pm \frac{|\vec{p}||\vec{\kappa}|}{m} \sqrt{ \cos^2(\theta_\kappa) \cos^2(\theta_p) + \sin^2(\theta_\kappa) \sin^2(\theta_p) }\, ,
\end{align}
with $\theta_p = \tan^{-1}(p_y/p_x)$ and $\theta_\kappa = \tan^{-1}(\kappa_y/\kappa_x)$.

The continuum eigenvalues are given by
\begin{align}
	\lambda_\pm &= -i \omega - \mu + \epsilon_\pm(\vec p, \vec \kappa).
\end{align}
Notice that, for $|\vec \kappa| = 0$, $\lambda_\pm$ vanishes at zero momentum and $\mu = 0$, signaling the well known instability
associated with Bose-Einstein condensation. For $|\vec \kappa| \neq 0$, the instability shifts to $\mu \neq 0$ and exists for both
$|\vec p| = 0$ and $|\vec p| \neq 0$.
The continuum eigenvalues have been studied in three dimensions in the Hamiltonian formulation for isotropic spin-orbit coupling in Ref.~\cite{Ozawa:2011kj}.
Additionally, one can see that $N_\pm = \varphi_\pm^\dagger \varphi_\pm$ are conserved, and that $N_+ + N_- = N_\uparrow + N_\downarrow$.

Using trigonometric identities it is possible to show that $v^+_i + v^-_i = \xi (1 - \cos(p_i) \cos(\kappa_i))/m$ and $v^+_i - v^-_i = \xi \sin(p_i) \sin(\kappa_i) / m$, from which it is clear that $\kappa_i=\pm\pi/2$ corresponds to the limit of $|\kappa_i| \gg |p_i|$ in the continuum.
When $|\vec{\kappa}| \gg |\vec{p}|$, such that the $|\vec{p}|^2$ term can be ignored, the single particle Hamiltonian becomes an anisotropic Weyl Hamiltonian with $\vec{v}_0 = \vec{\kappa}/m$ playing the role of the (anisotropic) speed of light and an effective chemical potential of $-m |\vec{v}_0|^2/2$.
A similar case has been discussed, in the context of ultracold fermionic atoms, in Ref.~\cite{2008PhRvA..77a1802J}.

\section{Many-body method}

The first order time derivative in the action is a non-Hermitian operator, making $e^{-S}$ a complex weight for the path integral; 
this is known as phase (or sign) problem.
This prevents the use of traditional Monte Carlo methods, since they use $e^{-S}$ as a probability weight.
One alternative in this scenario is the complex Langevin technique, which has been used to study theories with sign problems such as those with repulsive interactions~\cite{Loheac:2017yar} as well as polarized~\cite{Loheac:2018yjh, Rammelmuller:2018hnk} and mass-imbalanced fermions~\cite{Rammelmuller:2017vqn} (see Ref.~\cite{Berger:2019odf} for a review), finite density QCD with staggered quarks~\cite{Sexty:2013ica,Attanasio:2018rtq,Kogut:2019qmi,Sexty:2019vqx}, random matrix models~\cite{Nagata:2016alq,Bloch:2017sex}, rotating bosons~\cite{Hayata:2014kra,Berger:2018xwy}, superstring-inspired matrix models~\cite{Nishimura:2019qal}, among others.

The complex Langevin method is an extension of stochastic quantisation~\cite{Parisi:1980ys}.
The latter method consists of evolving the fields along a fictitious time dimension, $\theta$, according to the Langevin equation
\beq\label{eq.Langevin}
\frac{\partial \phi_s(x,\tau)}{\partial \theta} = -\frac{\delta S}{\delta \phi_s(x,\tau)} + \eta_s(x,\tau)\,,
\eeq
where $\eta_s(x,\tau)$ is a Gaussian white noise field satisfying
\begin{align}
\langle \eta_s (x,\tau) \rangle_\eta &= 0\,, \\
\langle \eta_s (x,\tau) \eta_{s'} (x',\tau') \rangle_\eta &= 2 \delta(x-x') \delta(\tau-\tau')\delta_{ss'}\,,
\end{align}
with $\langle \cdot \rangle_\eta$ indicating an ensemble average over the noise field. Quantum expectation values are obtained as
\beq
\langle \hat{O} \rangle = \lim_{\theta \to \infty} \langle \hat{O}(\phi_\uparrow(\theta),\phi_\downarrow(\theta)) \rangle_\eta \,,
\eeq
where $O$ is some observable. In practice, the Langevin equations are solved numerically with a step size $\varepsilon > 0$, chosen adaptively \cite{Aarts:2009dg}.
We follow an Euler-like discretization scheme in this work.
This generates a sequence of field configurations. Ensemble averages are performed as simple averages of the observables calculated using the configurations generated after the system reaches its steady state.

In order to deal with theories that have a complex action, each of the fields has to be complexified~\cite{Parisi:1984cs,Klauder:1983nn,Klauder:1983zm,Klauder:1983sp,Aarts:2008rr,Aarts:2008wh,Aarts:2009uq}.
For complex fields, both the real and imaginary parts become complex and obey the Langevin equation (\ref{eq.Langevin}). We choose the noise to remain real~\cite{Aarts:2011ax}. Expectation values are calculated in the same way as in the case with real action. Note that real observables do become complex in this method, but their imaginary parts should be statistically compatible with zero.

The average density was calculated as
\begin{align}
	\langle n \rangle &= \frac{1}{V} \frac{\partial \ln \mathcal Z}{\partial {(\beta \mu)}} = \frac{1}{V} \sum_{x,\tau} \langle n_{x,\tau} \rangle \nonumber\\
	&= \frac{e^{\xi \mu}}{V} \sum_{x,\tau} \left\langle \Phi_{(x,\tau)}^\dagger \Phi_{(x,\tau-a_\tau)} \right\rangle\,,
\end{align}
where the angular brackets on the right-hand side indicate an average over configurations generated by the Langevin process.
Mean-field results can be obtained by solving the Langevin equation without noise, which finds the minimum of the action.

We have performed our simulations on a periodic lattice of volume $20^2 \times 64$, a spacing anisotropy of $\xi=1/8$, and mass $m=1$.
These parameters lead to a thermal wavelength of $\sim 7$ in lattice units, which is consistent with the continuum limit window $1 \ll \lambda_T/a \ll N_x$.
The Langevin step size was chosen adaptively, with average of $\mathcal{O}(10^{-4})$.
We have estimated the auto-correlation time via the method proposed in~\cite{Wolff:2003sm}.

\section{Results}

\subsection{Quadratic case}
\begin{figure}
	\centering
	\includegraphics[width=0.95\linewidth]{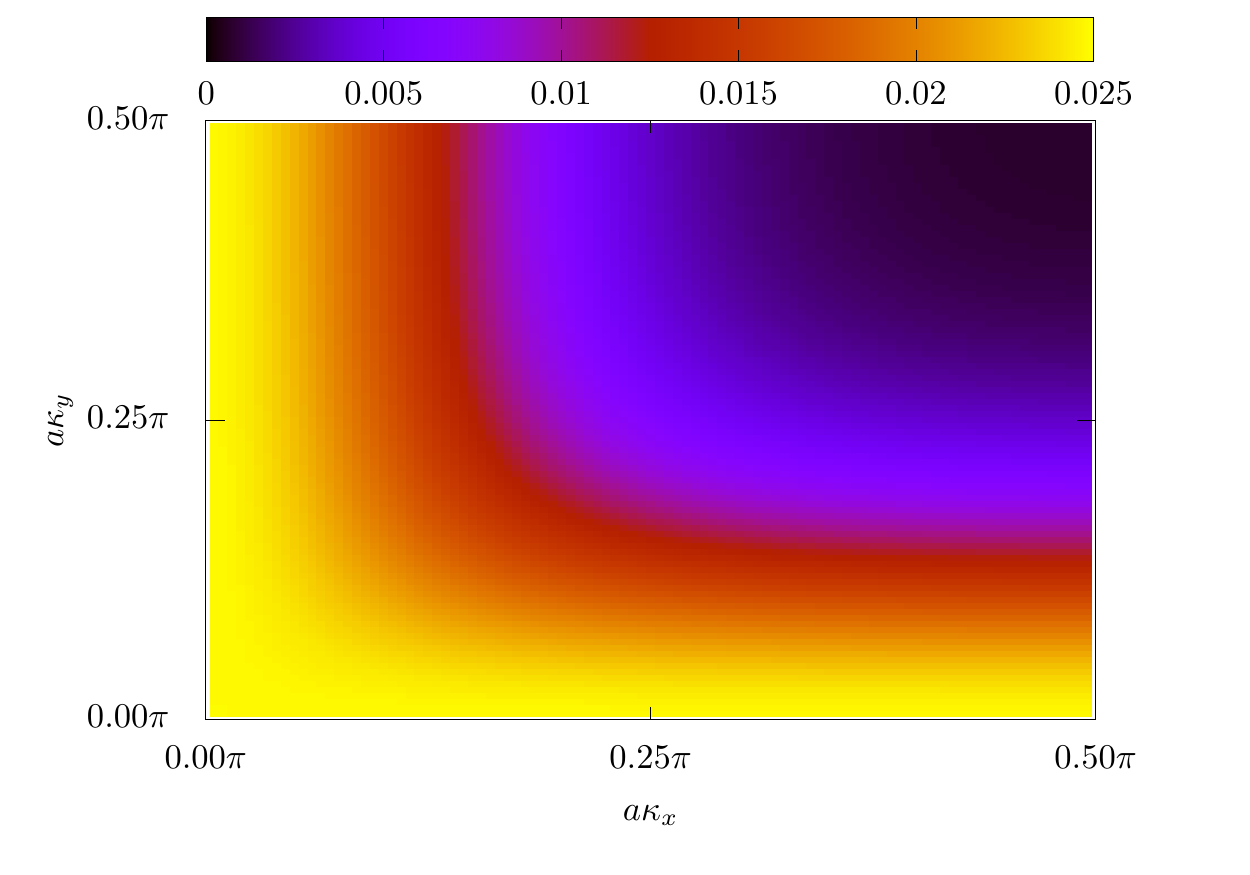} 
	\caption{\label{fig.density.quadratic} Average particle number density as a function of the spin-orbit couplings at $\lambda=0$. The simulations were performed at $\beta\mu=-0.8$.}
\end{figure}

We have studied the case of non-interacting ($\lambda=0$) exactly on the lattice.
The average particle number density can be seen in figure \ref{fig.density.quadratic}.
One can see that the density has its minimum when the system becomes Weyl-like, i.e., when the $p^2$ term in the Hamiltonian is much smaller than $\vec{\sigma} \cdot \vec{p}$ and can be neglected.

\begin{figure}
	\centering
	\includegraphics[width=0.95\linewidth]{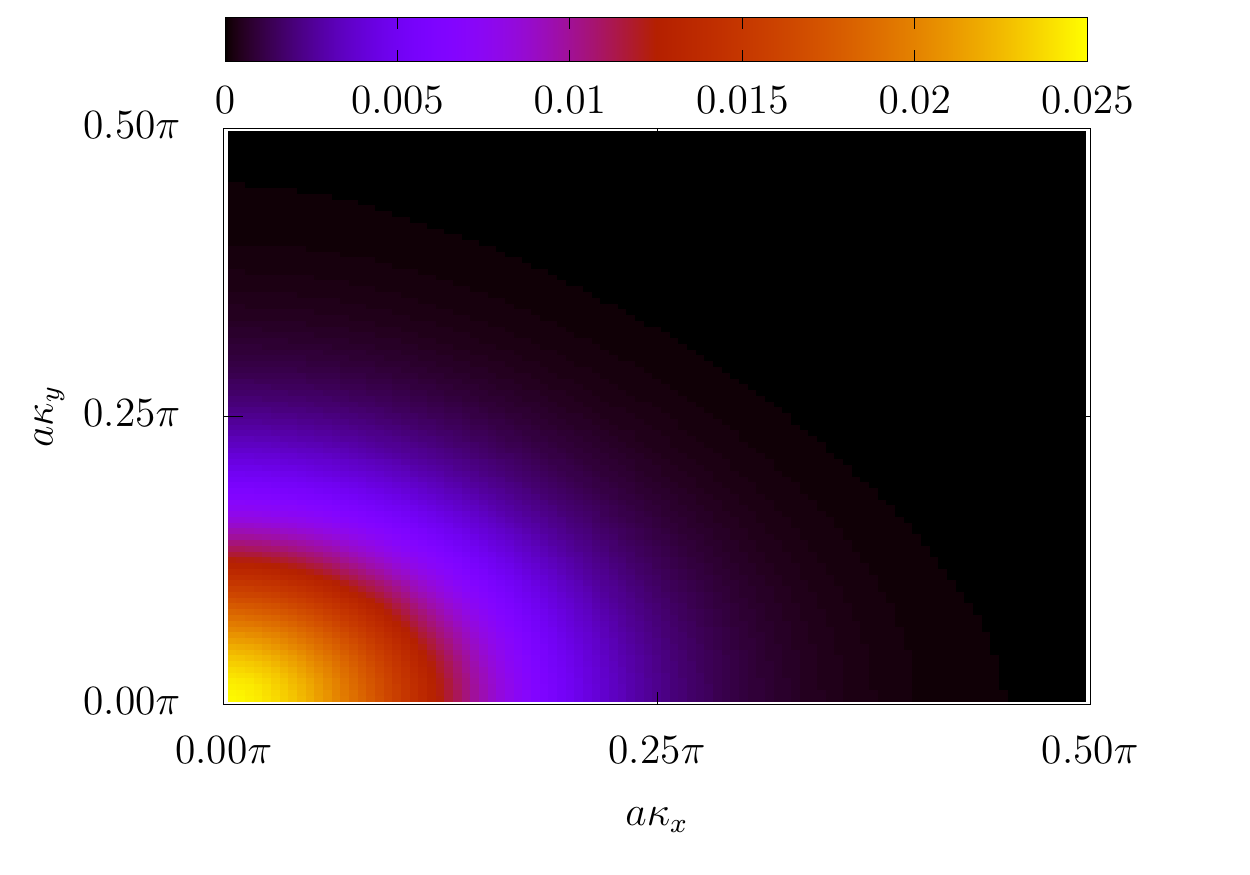} 
	\caption{\label{fig.density.quadratic.kappaMu} Average particle number density as a function of the spin-orbit couplings at $\lambda=0$ and ignoring the $\sin(\kappa_j)$ terms. The simulations were performed at $\beta\mu=-0.8$.}
\end{figure}

At $\beta \mu \geq 0$ and $\kappa_x = \kappa_y = 0$ the condensation of bosons in the ground state makes the simulation unstable.
This condensation is sharper at $T=0$ and softer at $T>0$, which is our case.
For non-zero spin-orbit couplings, however, the chemical potential where such condensation happens is pushed to larger values due to the $\kappa^2$ term behaving as an effective chemical potential.
To better visualise this, we have calculated the average density as a function of $\kappa_x$ and $\kappa_y$ by including only their effect on the chemical potential, i.e., the terms proportional to $\cos(\kappa_j)$, and ignoring the $\sin(\kappa_j)$ terms ($\vec{\sigma}\cdot\vec{p}$ in the continuum).
The result is shown in Fig.~\ref{fig.density.quadratic.kappaMu}.
The figure further shows that the $\vec{\sigma}\cdot\vec{p}$ term in the continuum action has a non-trivial effect on the average density.

\subsection{Interacting case I -- isotropic SOC}
\begin{figure}
	\centering
	\includegraphics[width=0.95\linewidth]{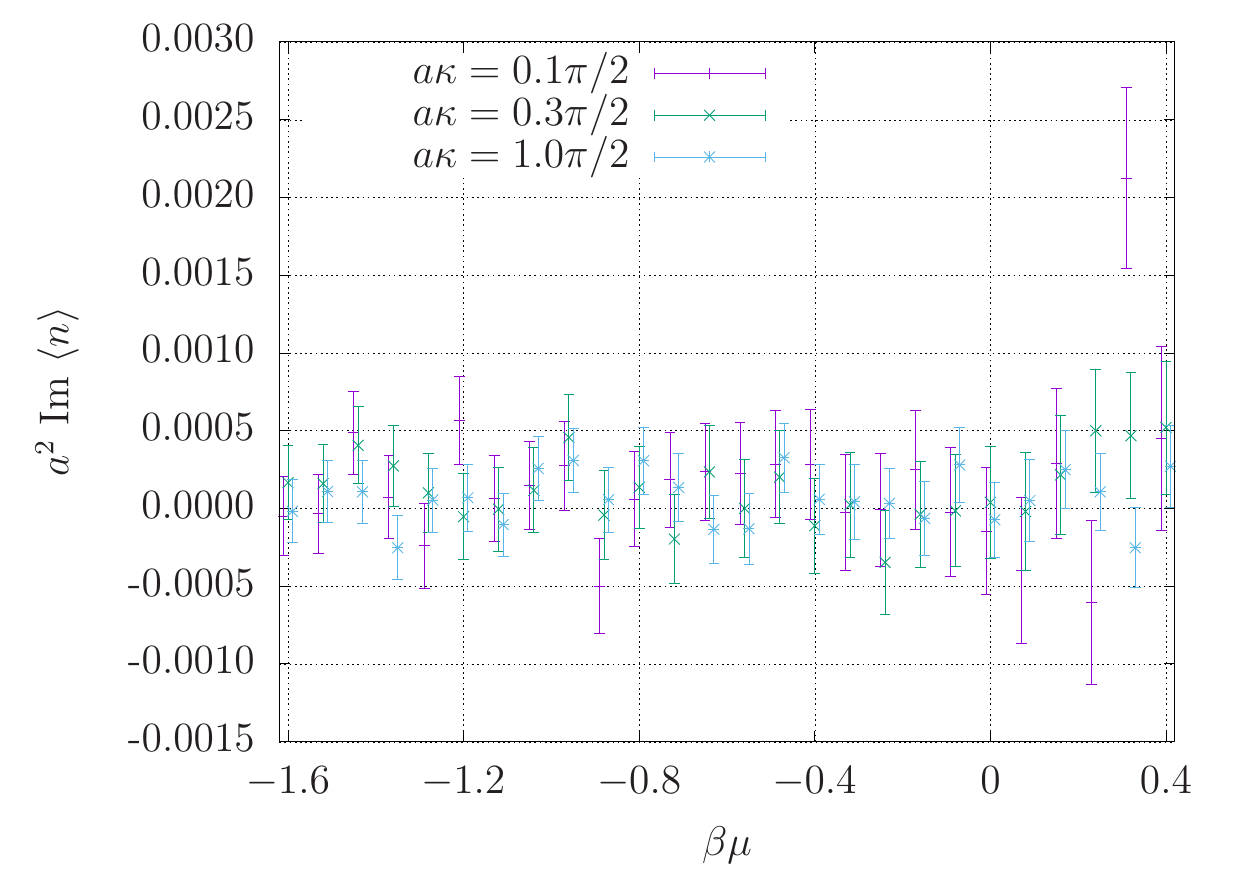}
	\caption{\label{fig.density.imag}Imaginary part of the average density as a function of $\beta\mu$ for different values of the spin orbit coupling. We have considered an interaction between particles of the same pseudo-spin, with coupling $\lambda/a=0.5$.
	Points have been slightly shifted horizontally for clarity.}
\end{figure}

As an initial check on the ability of the complex Langevin method to give correct results, we have looked at the imaginary part of the density.
Despite the sign problem in the Euclidean formulation, the density is expected to be (statistically compatible with) zero.
This can be verified in figure~\ref{fig.density.imag}.
Throughout this section we have used $\kappa_x = \kappa_y \equiv \kappa$.
Complex-valued densities would indicate a failure of the simulations.

\begin{figure}
	\centering
	\includegraphics[width=0.95\linewidth]{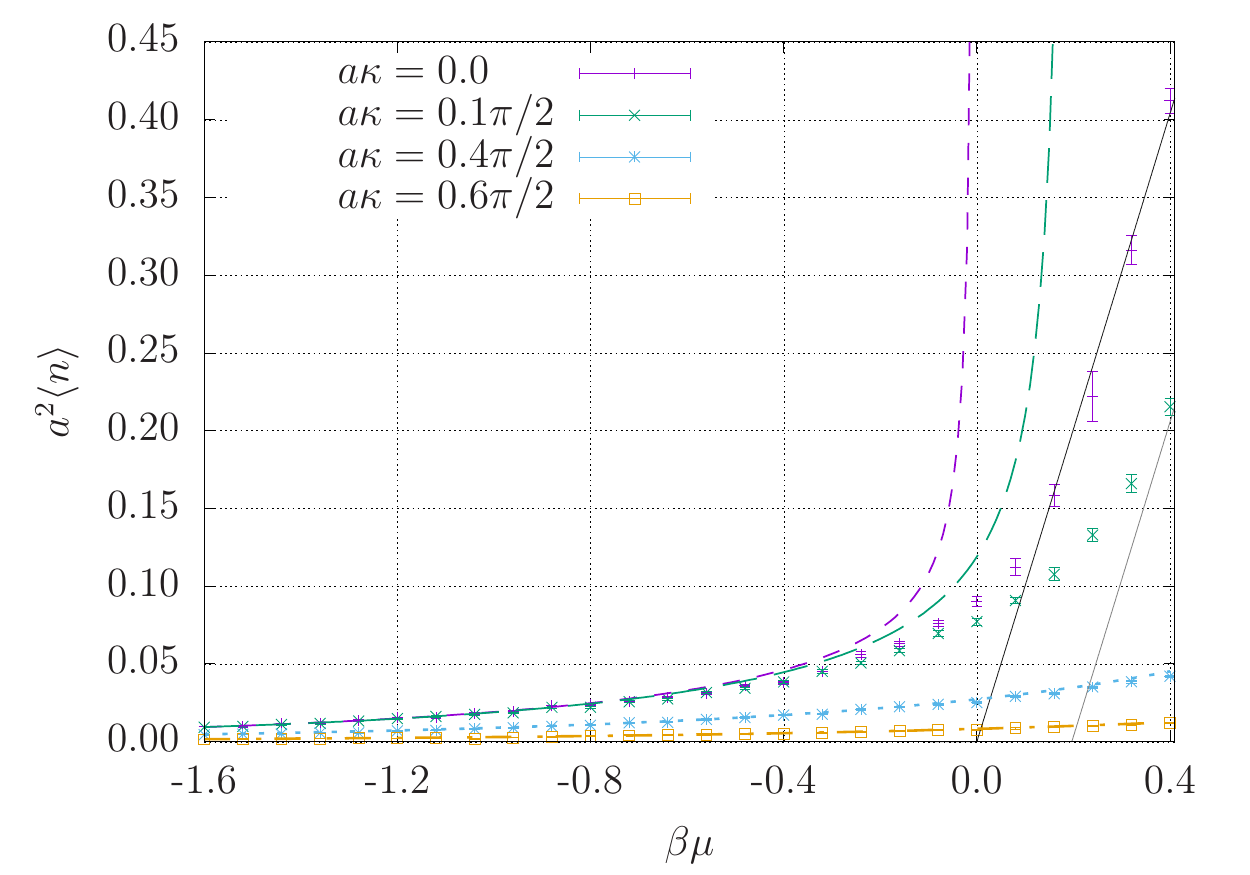}
	\caption{\label{fig.density.eos1} Average total density, for different values of $\kappa$ as a function of $\beta\mu$ at $\lambda/a=0.5$.
	We also show the non-interacting results for SOC of $\kappa=0$ (short dashes), $a\kappa=0.1\pi/2$ (long dashes), $a\kappa=0.4\pi/2$ (dotted), and $a\kappa=0.6\pi/2$ (dash-dotted).
	The continuous lines indicate the mean field result at $\kappa=0$ (darker) and $a\kappa=0.1 \pi/2$ (lighter).}
\end{figure}

The repulsive quartic couplings, similarly to the spin-orbit couplings, have the effect of keeping the system stable at small but positive chemical potentials.
This can be seen in Fig. \ref{fig.density.eos1}, where we show the average density as function of $\beta \mu$ for different values of (isotropic) SOC at $\lambda/a = 0.5$.
As in the case of $\lambda=0$, indicated by the long-dashed, short-dashed, dotted and dash-dotted lines for each $\kappa$, we observe smaller densities as the spin-orbit coupling increases.

The mean field average density is also shown in the figure.
For each flavour, it is given by
\beq
	\langle n_s \rangle = \frac{2 e^{\xi\mu}(e^{\xi\mu}-1-\xi/m(\cos(\kappa_x)+\cos(\kappa_y)-2))}{\xi \lambda}\,,
\eeq
and is zero when the right-hand side is negative.
For the larger values of $\kappa$ shown in the figure, the mean field density becomes positive at much higher values of $\beta\mu$.
The mean field average density for $a\kappa=0.4\pi/2$ and $0.6\pi/2$ is very small for the chemical potentials considered and not shown on the figure.
It is clear that the average density is not well described by the mean field result for $\mu \leq 0$ and/or $\kappa > 0$.

\begin{figure}
	\centering
	\includegraphics[width=0.95\linewidth]{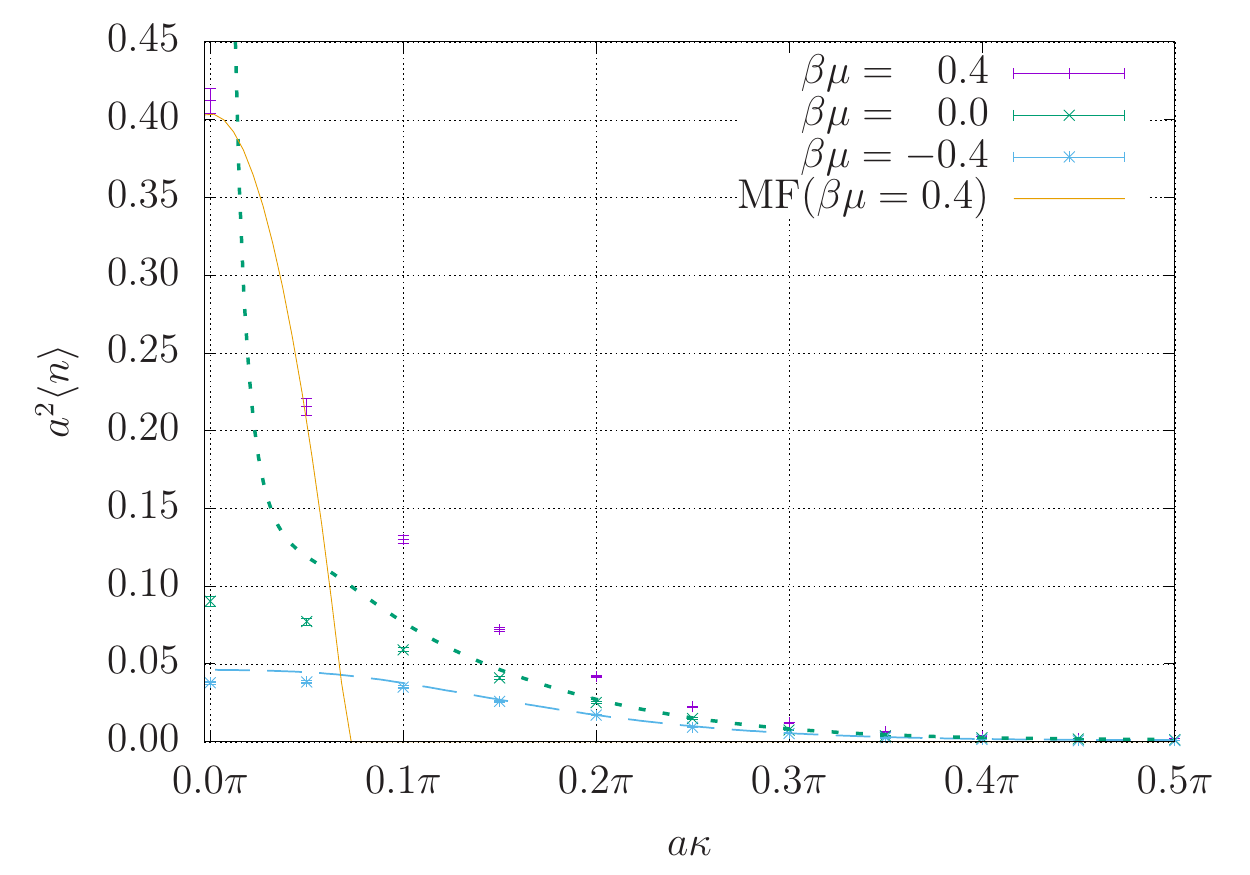}
	\caption{\label{fig.density.eos.kappa} Plot of the average total density as a function of $\kappa$ at $\lambda/a=0.5$. The solid line indicates the mean field result at $\beta\mu=0.4$, while the dashed line and dotted lines show the non-interacting results for $\beta\mu=-0.4$ and $\beta\mu=0.0$, respectively.}
\end{figure}

In order to have a better look at the effect of the SOC over the bosonic system, we show in Fig.~\ref{fig.density.eos.kappa} how the average density changes as we vary $\kappa$.
For comparison, we also show the mean field density at $\beta\mu=0.4$ and non-interacting results at $\beta\mu=-0.4$ and $\beta\mu=0.0$.
In all cases, an increase in the spin-orbit coupling has led to decreasing average densities as the system gets closer to the Weyl-like state.

\begin{figure}
	\centering
	\includegraphics[width=0.95\linewidth]{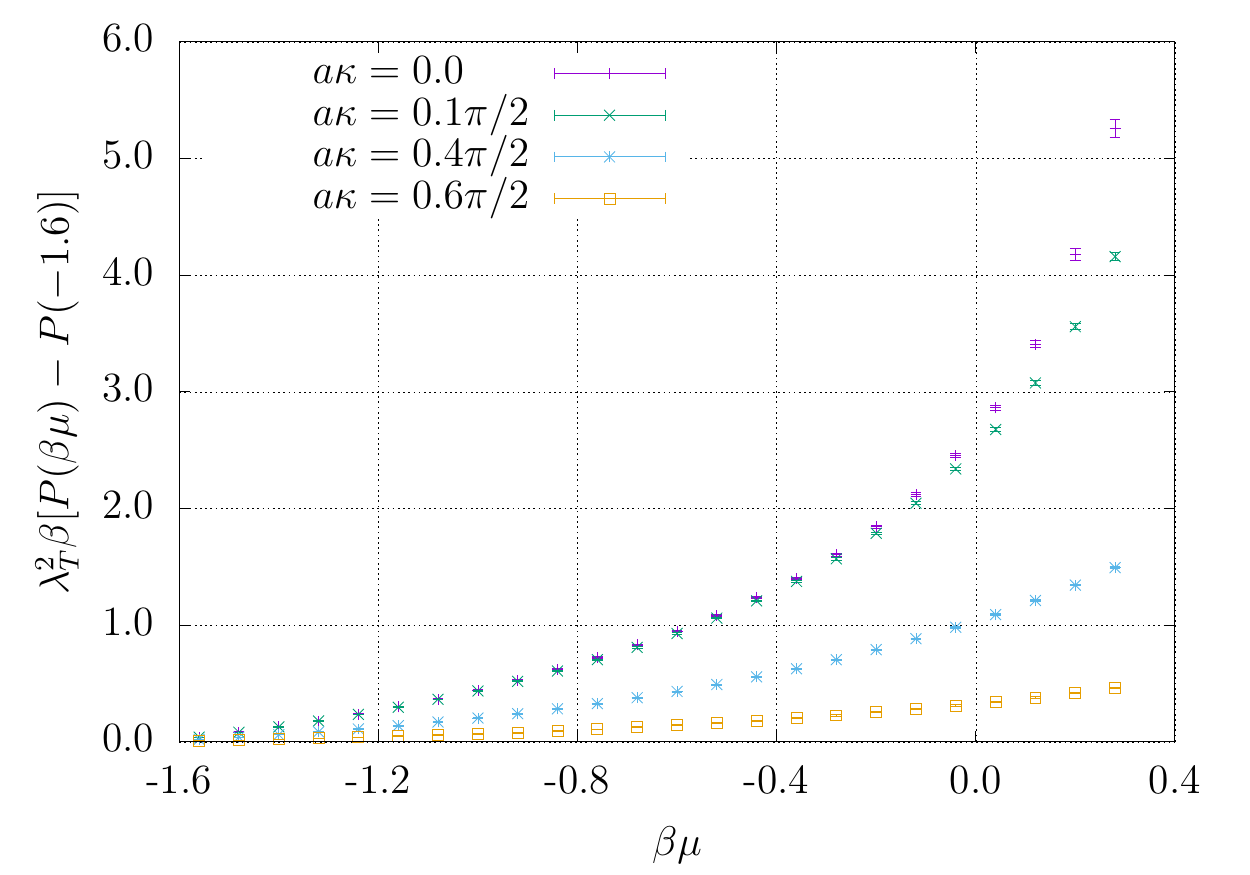}
	\caption{\label{fig.pressure.isotropic} Pressure difference as a function of $\beta\mu$ for different spin-orbit couplings.
	We have used the pressure at $\beta\mu_0=-1.6$ as reference for each SOC, and plotted in units of $(\beta\lambda_T^2)^{-1}$.}
\end{figure}

We have also calculated the pressure difference from a reference value via the Gibbs-Duhem relation
\beq
	P(\mu) - P(\mu_0) = \int_{\mu_0}^{\mu} \langle n(\mu') \rangle d\mu' \,.
\eeq
The numerical integration has been carried out using the trapezoid rule, with $\beta\mu_0 = -1.6$ as reference point.
We have estimated the errors via bootstrapping with $1000$ samples.
The results are shown in figure~\ref{fig.pressure.isotropic}.
The pressure difference in the plot is shown in units of $1 /\beta \lambda_T^2$.

\begin{figure}
	\centering
	\includegraphics[width=0.95\linewidth]{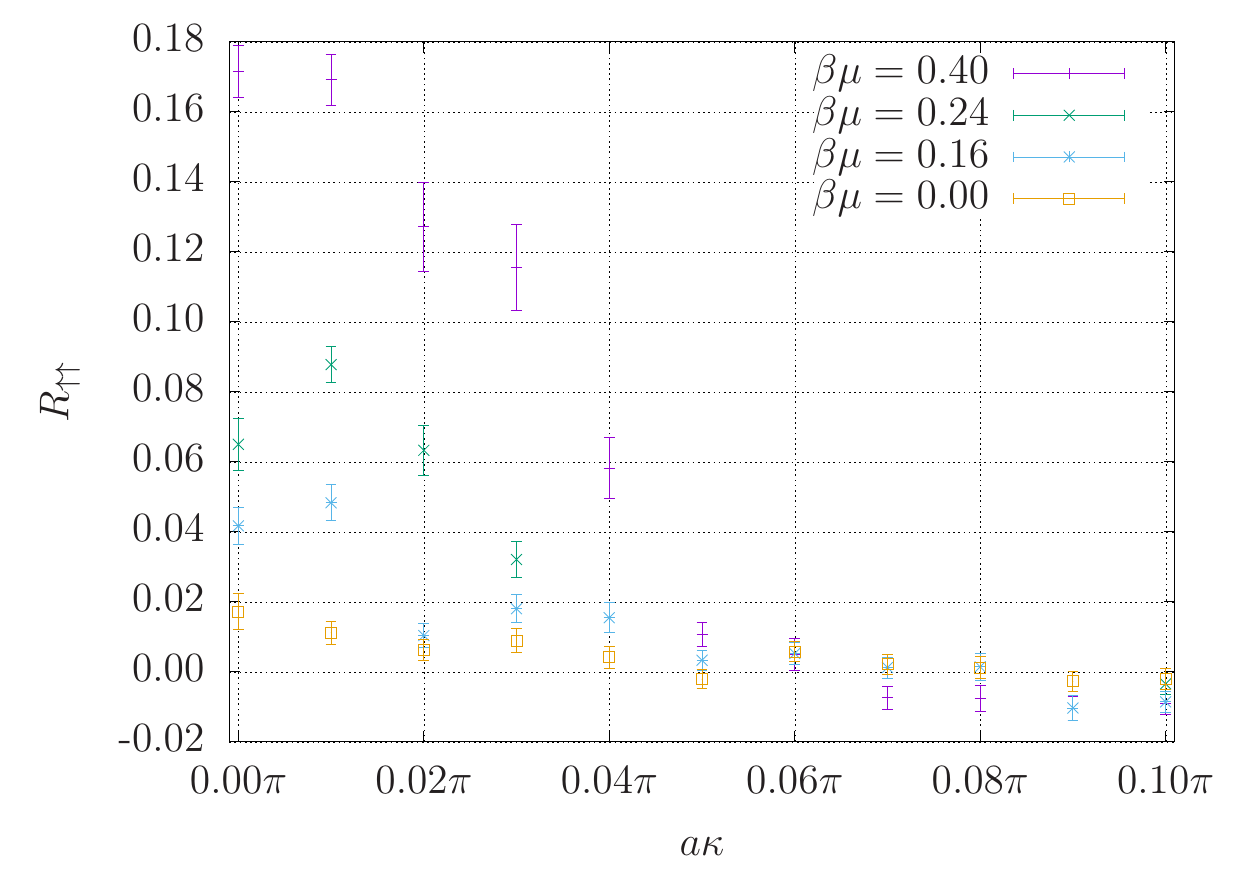}
	\caption{\label{fig.condensate.fraction} Pseudo-condensate fraction as function of the spin-orbit coupling for different chemical potentials.
	The figure shows how the off-diagonal long-range order is destroyed by increasing the SOC.}
\end{figure}

The confined nature of the simulation volume can induce the formation of (pseudo-)condensates at finite temperature.
These condensates manifest themselves as off-diagonal long range order in the correlation function between similar spins.
The correlation function is, in general, given by
\beq
	G_{ss'}\left(|\vec{x}-\vec{y}|\right) = \left\langle \phi^\dagger_{s(\vec{x},\tau)} \phi_{s'(\vec{y},\tau)}\right\rangle \,.
\eeq
We show, in figure~\ref{fig.condensate.fraction}, the condensate fraction,
\beq
	R_{ss'} = \frac{G_{ss'}(aN_x/2)}{G_{ss'}(0)}\,,
\eeq
between two pseudo-spin ``up'' fields, as a function of $\kappa$ and different values of $\beta\mu$.
A similar result is obtained for two ``down'' spins.
At $\kappa=0$ we observe a finite condensate fraction, which tends to zero as $\kappa$ increases, and is larger for higher values of $\beta\mu$.
Beyond $a\kappa>0.1\pi$, our results for $R$ are compatible with zero and thus excluded from the plot.
The correlations between fields of different spins have been measured to be statistically compatible with zero for all values of $\kappa$ and $\mu$ considered.

The above results indicate a destructive interplay between the spin-orbit coupling and condensation in finite systems in two spatial dimensions.
A similar phenomenon has been observed in three dimensions with an inter-species coupling in~\cite{2018PhRvL.120n0403L}, and isotropic $s$-wave coupling~\cite{PhysRevLett.109.025301}.

\subsection{Interacting case II -- anisotropic SOC}
We have investigated the effects of anisotropic spin-orbit coupling, i.e., $\kappa_x \neq \kappa_y$, on the density equation of state by using $\kappa_y = \eta_{\mathrm{soc}} \kappa_x$, with $0 \leq \eta_{\mathrm{soc}} \leq 1$.
Because of the symmetry $\kappa_x \leftrightarrow \kappa_y$ in the partition function there is no need to consider $\eta_\mathrm{soc} > 1$.

\begin{figure}
	\centering
	\includegraphics[width=0.95\linewidth]{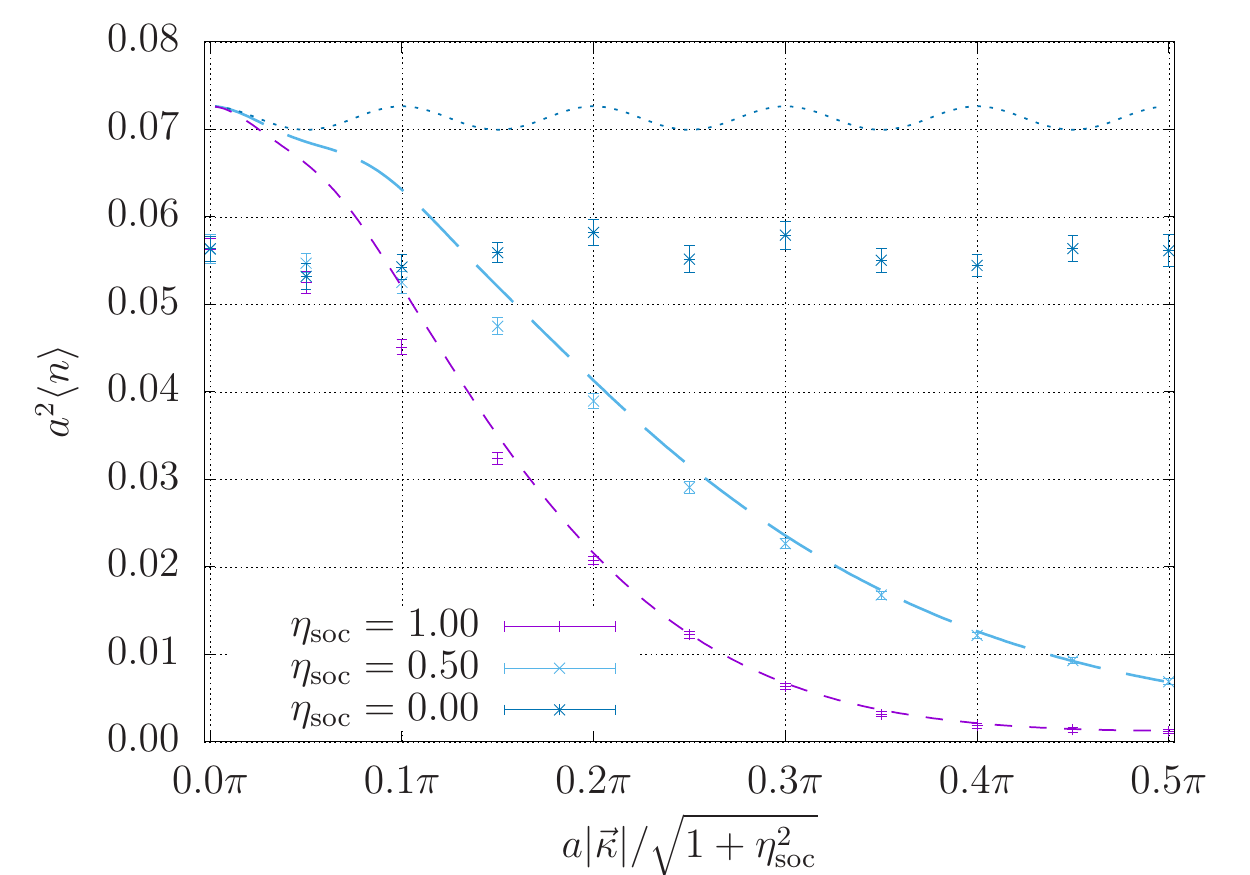}
	\caption{\label{fig.density.kappa.aniso} Average number density as function of the absolute value of the spin-orbit coupling vector for $\beta\mu = -0.2$.
	The lines display the results for the non-interacting case of $\lambda=0$: short dashes shows $\eta_{\mathrm{soc}}=1.0$, long dashes stand for $\eta_{\mathrm{soc}}=0.5$, and the dotted line represents $\eta_{\mathrm{soc}}=0.0$.}
\end{figure}

In figure~\ref{fig.density.kappa.aniso} we plot the average density as function of $|\vec{\kappa}|$.
The $x$-axis has been normalized so that the maximum value of $|\vec{\kappa}|$ is the same for all anisotropies.
We observe a slower decay of $\langle n \rangle$ as a function of $|\vec{\kappa}|$ for $\eta_{\mathrm{soc}} < 1$.
We remind the reader that the mean field densities for negative chemical potentials are zero.
At $\eta_{\mathrm{soc}}=0.0$ there is a periodic behaviour of period $2\pi/N_x$ on the average density for both non-interacting and interacting cases.
In the former, we have verified it to be a property of the partition function as a whole, and not of the eigenvalues.
Some remnant of this periodicity can be seen at $\eta_{\mathrm{soc}}=0.5$ for small values of $|\vec{\kappa}|$.

\begin{figure}
	\centering
	\includegraphics[width=0.95\linewidth]{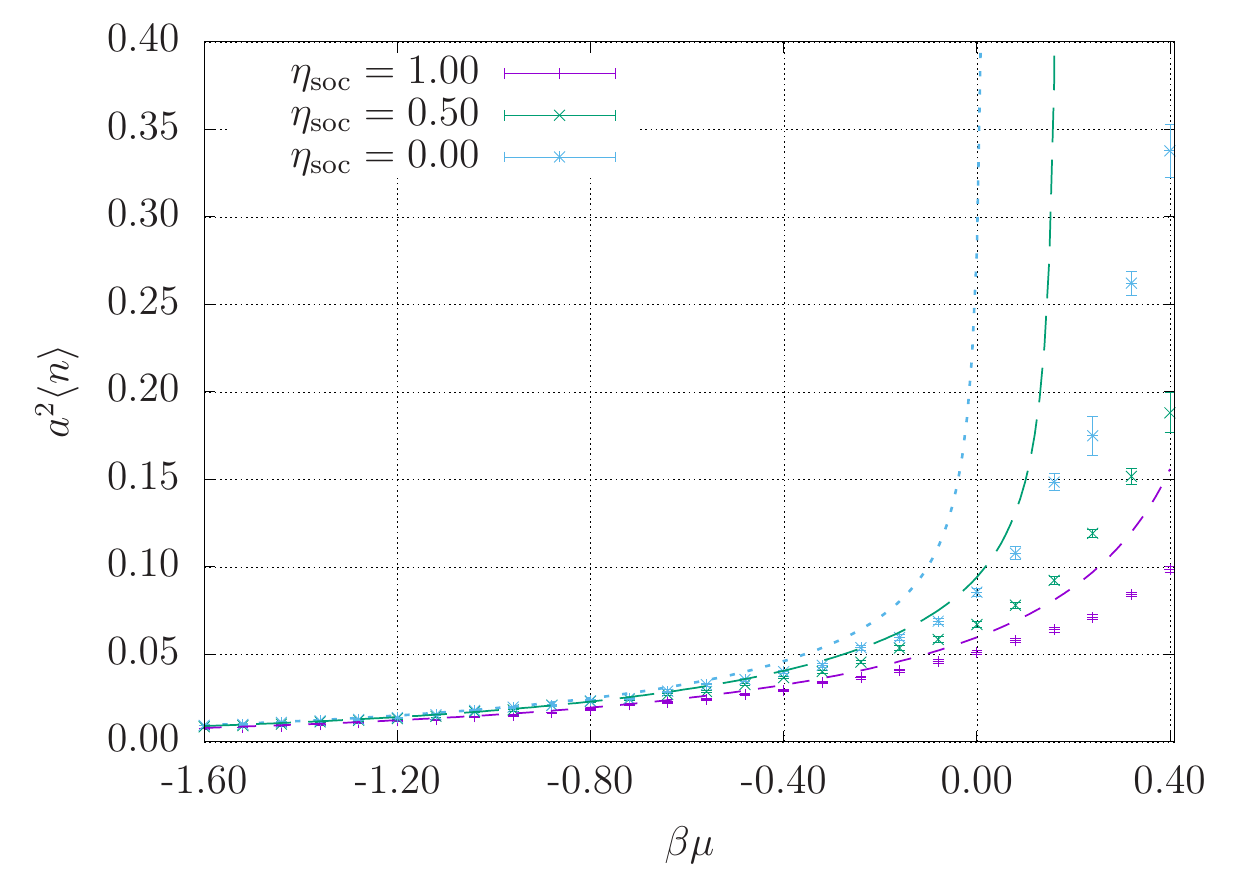}
	\caption{\label{fig.density.eos.aniso2} Density equation of state for anisotropic spin-orbit couplings at $a\kappa_x=0.25\pi/2$ and $\lambda/a = 0.5$.
	Also shown are the non-interacting results for $\eta_{\mathrm{soc}}=1.0$ (short dashed line), $\eta_{\mathrm{soc}}=0.5$ (long dashed line), and $\eta_{\mathrm{soc}}=0.0$ (dotted line).
	Mean field results are very small and were omitted.}
\end{figure}

The density equation of state for different SOC anisotropies is shown in figure~\ref{fig.density.eos.aniso2} for both interacting (points) and non-interacting (lines) cases.
We have used $a\kappa_x=0.25\pi/2$.
Mean field results are very small in comparison to those in the figure and therefore omitted.
The figure shows that the distincion between the interacting and non-interacting equation of states becomes more pronounced as $\eta_{\mathrm{soc}}$ decreases, and is stronger for larger chemical potentials.

\begin{figure}
	\centering
	\includegraphics[width=0.95\linewidth]{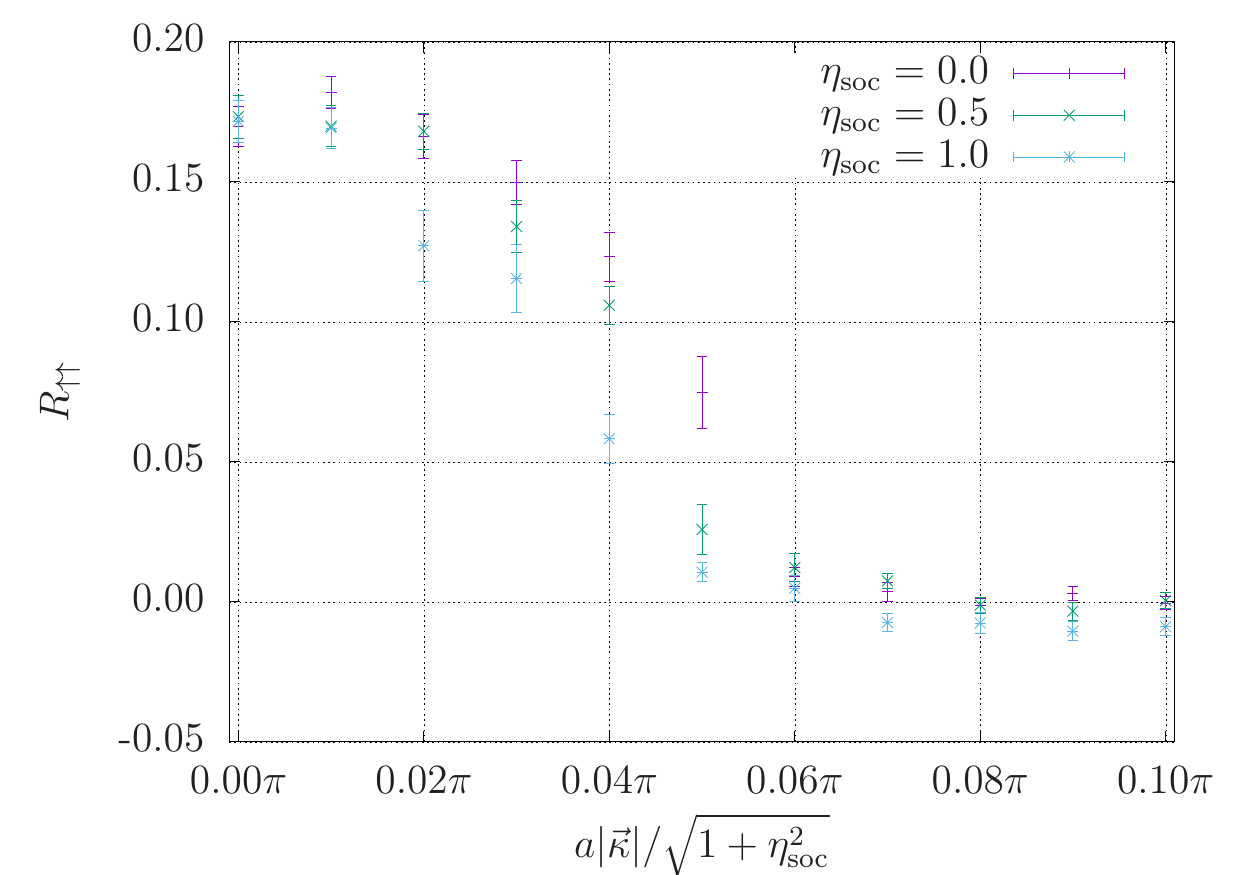}
	\caption{\label{fig.condensate.fraction.aniso} Plot of the pseudo-condensate fraction as function of the absolute value of the spin-orbit coupling magnitude, for different SOC anisotropies at $\beta\mu=0.4$.
	The off-diagonal long-range order survives longer for smaller values of $\eta_{\mathrm{soc}}$.}
\end{figure}

As with the isotropic spin-orbit case, we have investigated the condensate fractions for $\eta_{\mathrm{soc}}<1$.
Figure~\ref{fig.condensate.fraction.aniso} shows the results for $R_{\uparrow\uparrow}$ as a function of $|\vec{\kappa}|$.
Similar to the $\eta_{\mathrm{soc}}=1$ case, $R_{\downarrow\downarrow}$ displayed a similar behaviour.
We observe a slower decay of both of the (pseudo)-condensate fraction for larger SOC anisotropies.
This effect has been seen in three dimensions in~\cite{PhysRevA.87.031604}.

\section{Summary and outlook}
We have investigated the effects of an artificial spin-orbit coupling on the density equation of state for a bosonic system of two pseudo-spins, as well as contact interactions between similar boson species using non-perturbative numerical methods.
The action is complex in Euclidean spacetime due to the first order time derivative, and therefore our method of choice for the simulations was the complex Langevin technique.

Due to how the spin-orbit term enters the lattice formulation, it was possible to investigate how the equation of state changes in the range $0 \leq \kappa_x , \kappa_y < \infty$.
In particular, the Hamiltonian becomes Weyl-like when the spin-orbit coupling is much larger than the momentum.

We have obtained the density and pressure equations of state for different values of the spin-orbit coupling in the isotropic case, where $\kappa_x = \kappa_y \equiv \kappa$.
The average number density has been seen to be a decreasing function of $\kappa$, having its minimum value when $a\kappa=\pi/2$ (which corresponds to $\kappa \to \infty$ in the continuum).
A comparison with mean field results has shown that quantum effects play a bigger role for larger values of $\kappa$ and positive chemical potentials.
For $\mu \leq 0$ the mean field average density is zero, in clear contrast with the simulations, which include all quantum effects.

We have also investigated the case of anisotropic spin-orbit coupling, with $\kappa_y = \eta_{\mathrm{soc}} \kappa_x$.
A periodic behaviour of the number density as a function of the spin-orbit coupling, induced by the finite volume, has been observed in both interacting and non-interacting cases when the SOC anisotropy $\eta_{\mathrm{soc}}=0$.
As $\eta_{\mathrm{soc}}$ is increased from $0$ towards $1$, the aforementioned decaying behaviour of the density is recovered.

For both isotropic and anisotropic spin-orbit couplings, we have carried out comparisons between the interacting and non-interacting equations of state.
The distinction between them is smallest for large SOC, and becomes more apparent as $|\vec{\kappa}| \to 0$.
Moreover, the finite volume of the simulations allows the formation of a pseudo-condensate, which is depleted by the spin-orbit coupling.

Possible future studies include non-perturbative investigations of the interplay between the spin-orbit coupling and rotation, as well as different types of SOC~\cite{PhysRevLett.107.150403}.
Moreover, the determination of physical quantities such as the scattering length or binding energy, which help connecting with experimental results, can be done via L\"{u}scher's method~\cite{luscher1986} or the second virial coefficient~\cite{Shill:2018tan}.

\section{Acknowledgements}
	The authors would like to thank T. Ozawa for useful discussions.
	This work was facilitated though the use of advanced computational, storage, and networking infrastructure provided by the Hyak supercomputer system at the University of Washington.
	The work of FA was supported by US DOE Grant No. DE-FG02-97ER-41014.
	This material is based upon work supported by the National Science Foundation under Grant No.
PHY{1452635} (Computational Physics Program).
F.A and acknowledges the hospitality of the University of North Carolina at Chapel Hill where part of this work was performed.
\appendix

\bibliographystyle{style}
\bibliography{ref}
\end{document}